\documentclass[
  aps,
  pra,
  reprint,
  superscriptaddress,
  amsmath,
  amssymb,
  longbibliography
]{revtex4-2}

\usepackage[T1]{fontenc}
\usepackage[english]{babel}
\makeatletter
\let\l@en\l@english
\makeatother
\usepackage{graphicx}
\usepackage{braket}
\usepackage{mathrsfs}
\usepackage{bm}
\usepackage{mathtools}

\usepackage{hyperref}
\usepackage{balance}

\usepackage{xcolor}
\usepackage{float}

\usepackage[caption=false]{subfig}

\graphicspath{{figures/}}

\begin{document}

\title{Floquet-Engineered Parity Anomaly Staircase in a Cold Atom Dirac Lattice}

\author{Binayyak Roy}
\affiliation{Department of Physics and Astronomy, Clemson University, Clemson, South Carolina 29634, USA}

\author{Vito Scarola}
\affiliation{Department of Physics, Virginia Tech, Blacksburg, Virginia 24061, USA}

\author{Sumanta Tewari}
\affiliation{Department of Physics and Astronomy, Clemson University, Clemson, South Carolina 29634, USA}

\date{\today}

\begin{abstract}
We propose a Floquet-engineered cold atom realization of a parity anomaly inspired anomalous Hall staircase in a two dimensional $\pi$-flux lattice. The effective model hosts massive Dirac fermions generated by the combined action of a time reversal symmetry breaking Floquet mass and a static inversion breaking mass offset. An additional momentum dependent scalar displacement term shifts different Dirac sectors in opposite energy directions without modifying their Bloch eigenvectors. As a result, the Berry curvature contribution associated with individual massive Dirac sectors can be selectively occupied, allowing the anomalous Hall response to evolve stepwise as a function of chemical potential or scalar displacement term. Evaluating the full lattice Berry curvature integral, we find plateau-like responses near $0$, $e^2/2h$, and $e^2/h$, corresponding respectively to the activation of zero, one, and two effective massive Dirac sector contributions. We analyze the associated low energy Dirac theory, band topology, Berry curvature structure, and two parameter response maps, and discuss a possible realization using Raman-assisted tunneling, off-resonant Floquet driving, and auxiliary AC-Stark dressing in ultracold atomic optical lattices.
\end{abstract}

\maketitle

\section{Introduction}
\label{sec:introduction}

Quantum anomalies provide striking examples of how symmetries that appear valid at the classical level can fail after quantization. Among the most important examples in two spatial dimensions is the parity anomaly of a Dirac fermion. A massless Dirac Hamiltonian can preserve parity and time-reversal symmetries, while the introduction of a mass term generates a Hall response with half-integer contribution
\[
    \sigma_{xy}
    = 
    \frac{1}{2}\operatorname{sgn}(M)\frac{e^2}{h},
\]
where the sign is determined by the Dirac mass and chirality. This half-quantized contribution is a characteristic manifestation of parity-anomalous Dirac physics and is closely related to the anomalous Hall physics of massive Dirac fermions in Chern-insulator models \cite{NiemiPRL1983,CallanNUCLPB1985,Redlich1984,Jackiw1984,haldane_model_1988,QiPRB2008}.

In a microscopic lattice system, however, an isolated Dirac cone cannot generically appear under the usual assumptions of locality, Hermiticity, and translational invariance \cite{Nielsen1981}. Dirac cones instead occur in pairs of opposite chirality, so that the Hall response of completely filled bands remains integer quantized \cite{NovoselovNat2005,DengScience2020,klembt2018exciton,ChangAAAS2013,YuAAAS2010,SerlinAAAS2020,wang2009Nature}. As a result, the direct observation of half-integer Dirac contributions is subtle: the response associated with one massive Dirac cone is generally accompanied by additional contributions from other regions of the Brillouin zone. An important challenge is therefore to engineer systems in which the Berry curvature response associated with selected massive Dirac sectors can be energetically isolated and probed independently.

Cold atom platforms provide a natural setting for this program because optical lattices \cite{Dauphin2013APS,PricePRA2012,Goldman2014}, Raman assisted tunneling \cite{AidelsburgerPRL2011}, synthetic gauge fields \cite{polini2013artificial,dalibard2011colloquium,zhai2012spin,vozmediano2010gauge,guinea2010energy,levy2010strain,JosePRL2012}, and Floquet engineering allow single particle Hamiltonians to be constructed with a high degree of tunability \cite{Goldman2014,cooper_topological_2019,eckardt_colloquium_2017}. Recent work employing synthetic dimensions and coupled-wire geometries has demonstrated parity anomaly related half-quantized anomalous Hall responses near a topological phase transition critical point in ultracold atoms \cite{mittal_two-dimensional_2026}. Related studies have also shown that non-Abelian optical lattices can display half-odd topological number sequences associated with anomalous Landau level physics of Dirac fermions \cite{Mei2012}. These developments motivate the question, can one engineer a two dimensional optical-lattice system in which distinct massive Dirac sectors are shifted in energy and occupied sequentially, thereby producing an anomalous Hall response staircase with intermediate half-quantized plateaus, revealing the effects of parity anomaly on Dirac fermions?

In this work we propose an effective cold atom realization of such a parity anomaly inspired Hall staircase. We consider a two band $\pi$-flux optical lattice \cite{JiaPRB2013,OtsukaPRB2002,otsuka2014mott,li2015fermion,ParisenPRB2015,GuoPRB2018,zhang_charge_2020} whose kinetic Hamiltonian hosts four Dirac cones. Raman-assisted tunneling in a tilted optical lattice generates the $\pi$-flux hopping structure through laser induced Peierls phases \cite{aidelsburger_realization_2013,miyake_realizing_2013}, while an off-resonant circular Floquet drive induces a time reversal symmetry breaking mass \cite{Oka2009,Lindner2011}. A static inversion breaking offset provides an additional mass contribution, together producing Berry curvature localized near the gapped Dirac cones. In addition, we introduce an auxiliary scalar AC-Stark dressing channel that generates a momentum dependent scalar shift. Because this term multiplies the identity matrix, it shifts the energies of the effective Dirac sectors without modifying their Bloch eigenvectors.

\begin{figure*}[t]
    \centering
    \includegraphics[width=\linewidth]{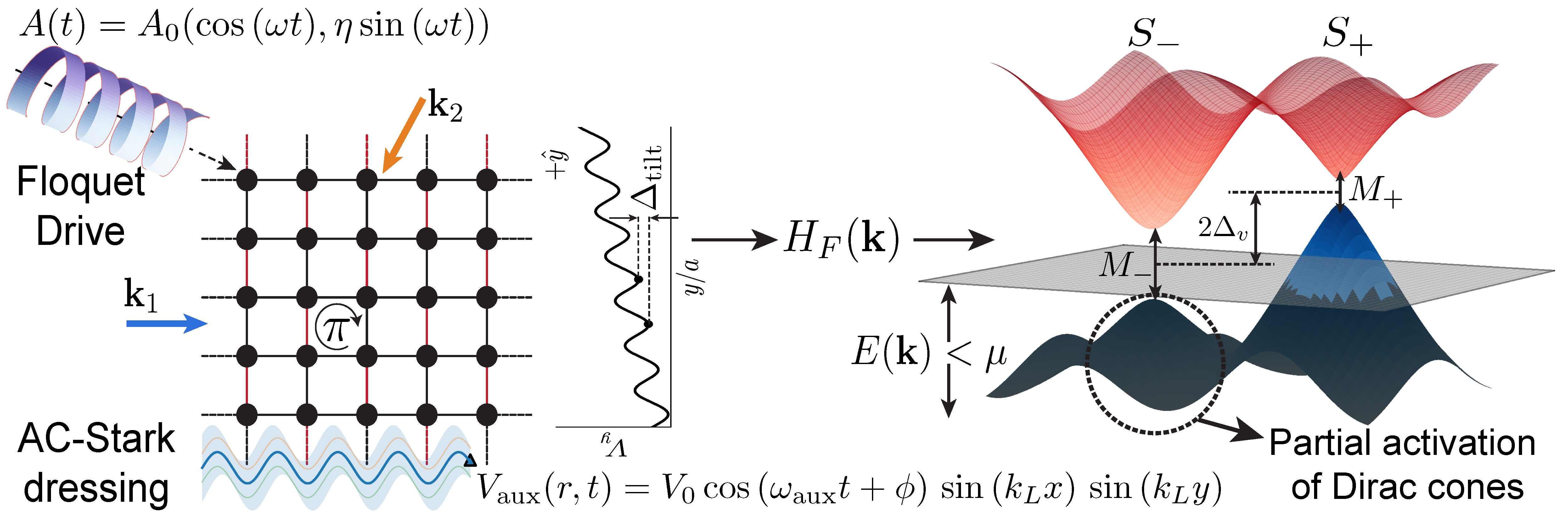}
    \caption{
    Schematic of the proposed Floquet-engineered $\pi$-flux optical lattice and the resulting Dirac sector activation mechanism. A static tilt suppresses bare tunnelling along the $\hat{y}$ direction, while Raman-assisted tunnelling restores resonant hopping with controlled Peierls phases such that the accumulated phase around a plaquette is $\pi$. The blue and yellow arrows indicate the Raman beams with wave vectors $\mathbf{k}_1$ and $\mathbf{k}_2$. Floquet drive $\mathbf A(t)=A_0(\cos\omega t,\eta\sin\omega t)$, shown by the helical driving field, generates a time reversal symmetry breaking Dirac mass. An additional auxiliary scalar AC-Stark dressing channel, $V_{\mathrm{aux}}(\mathbf r,t)$, shown by the wavy modulation field, produces a momentum dependent scalar quasienergy shift that separates the $S_+$ and $S_-$ Dirac sectors (refer to main text) in energy. The gray plane denotes the chemical potential $\mu$, and the dark blue regions indicate occupied states contributing to the Berry curvature integral. When $\mu$ intersects only a subset of the shifted massive Dirac cones, the system exhibits a partial Hall response arising from selective Dirac sector activation.
    }
    \label{fig:proposal_schematic}
\end{figure*}

The central mechanism is the separation between Berry curvature generation and Berry curvature occupation. The mass terms determine the local Berry curvature near the gapped Dirac cones, whereas the scalar shift controls the relative energies of the two effective sectors. Consequently, varying either the chemical potential $\mu$ or the scalar shift $\Delta_v$ produces a stepwise evolution of the occupied Berry curvature contribution. Using the full lattice Berry curvature integral, we find extended plateau-like regions near $\sigma_{xy}\simeq e^2/2h$ and $\sigma_{xy}\simeq e^2/h$. The intermediate response occurs when the occupied states are dominated primarily by Berry curvature associated with one effective massive Dirac sector, whereas the larger response appears when both relevant sectors contribute significantly. 

Although the present work focuses on the anomalous Hall response obtained from the Berry curvature integral, the corresponding transport signatures can be probed experimentally through semi-classical wavepacket dynamics and center of mass motion in optical lattices\cite{jotzu_experimental_2014, Aidelsburger_2014, Wintersperger_2020}. In the presence of an applied force $\mathbf{F}$, the crystal momentum evolves according to,

\begin{equation}
    \hbar\dot{\mathbf{k}} = \mathbf{F}
\end{equation}
so that, for example,
\begin{equation}
    k_x(t) = k_x(0) + \frac{F_x}{\hbar}t
\end{equation}
The semi-classical velocity then acquires an anomalous contribution proportional to the Berry curvature,
\begin{equation}
    \dot{\mathbf{r}} = \frac{1}{\hbar}\nabla_{\mathbf{k}}E_n(\mathbf{k}) - \dot{\mathbf{k}}\times\Omega_n(\mathbf{k})
\end{equation}
leading to a transverse drift of the atomic cloud under an applied force. Measurements of the transverse center of mass velocity have previously been used to extract Hall and Berry curvature responses in topological optical lattices \cite{PricePRB2016,Genkina_2019,Anderson_2019}. Such measurements provide an experimentally accessible probe of the occupied Berry-curvature distribution. In this sense, the staircase response discussed here may therefore be probed through the force-induced transverse center of mass dynamics of the occupied bands. Recent synthetic dimension experiments \cite{mittal_two-dimensional_2026} have also inferred parity anomaly related responses using internal atomic states as an additional synthetic coordinate. In contrast, this work is based on a fully spatial two dimensional Floquet-engineered $\pi-$flux lattice with selective scalar displacements of massive Dirac cones.

The paper is organized as follows. In Sec.~\ref{sec:model}, we introduce the effective Floquet engineered $\pi$-flux Hamiltonian and discuss the proposed cold atom implementation. In Sec.~\ref{sec:dirac_structure}, we derive the low energy Dirac expansion and identify the two effective sectors $S_+$ and $S_-$. In Sec.~\ref{sec:band_berry}, we analyze the band structure, Berry curvature, and cone resolved contributions to the Hall response. In Secs.~\ref{sec:mu_staircase} and \ref{sec:delta_staircase}, we examine the staircase response generated by chemical potential and scalar shift sweeps, respectively. Finally, in Sec.~\ref{sec:phase_diagrams}, we map the Hall response in the two parameter $(\mu,\Delta_v)$ plane and identify the parameter regimes in which the intermediate plateau-like responses remain robust.

\section{Model Hamiltonian}
\label{sec:model}

We consider a two dimensional two band lattice model of fermions describing in an effective Floquet-engineered $\pi$-flux optical lattice. The proposed mechanism is summarized schematically in Fig.~\ref{fig:proposal_schematic}. A static potential gradient $V_y$ produces an energy offset $\Delta_{\mathrm{tilt}}$ between neighboring sites along the $\hat{y}$ direction, suppressing bare tunneling on the tilted bonds. Resonant hopping is then restored by Raman-assisted tunneling which provides a standard route for engineering synthetic gauge fields and laser-induced Peierls hopping phases. The Peierls phases can be chosen such that the net phase accumulated around each plaquette is $\pi$ \cite{AidelsburgerPRL2011,miyake_realizing_2013,aidelsburger_realization_2013,Aidelsburger_2014}. This provides an experimentally motivated route to the $\pi$-flux hopping structure used below.

We further subject the effective lattice model to an off-resonant circular drive,
\begin{equation}
    \mathbf{A}(t)
    =
    A_0(\cos\omega t,\eta\sin\omega t),
\end{equation}
shown schematically by the helical field in Fig.~\ref{fig:proposal_schematic}. In the high frequency regime, such a periodic drive generates an effective time reversal symmetry breaking mass through the leading commutator term in the Floquet expansion \cite{Goldman2014,eckardt_colloquium_2017,mikami_brillouin-wigner_2016,KitagawaPRB2011}. We denote the corresponding mass scale by $m_T$. In addition, we include a static inversion breaking mass offset $m_I$, which may be implemented microscopically through a sublattice energy imbalance, differential light shift, or Raman detuning in the two-component $\pi$-flux unit cell. In the effective model $m_I$ acts as a momentum independent contribution to the Dirac mass thus breaking the inversion symmetry, while the circular Floquet drive produces the momentum dependent contribution $2m_T\sin k_x\sin k_y$ making it the analogue of Haldane mass \cite{haldane_model_1988} and reflecting time reversal symmetry breaking.

Finally, we introduce a momentum dependent scalar term $\Delta_v\sin k_x\sin k_y$. This term is intended to represent an auxiliary scalar dressing channel, for example an off-resonant AC-Stark or Raman dressing process that couples approximately symmetrically to the two pseudospin components \cite{SpielmanPRA2009,cooper_topological_2019, grimm_optical_2000}. Since it multiplies the identity matrix, it shifts both bands equally at fixed momentum and does not modify the Bloch eigenvectors. In this sense, the scalar term should be regarded as an effective quasienergy displacement of the Dirac sectors, rather than as an additional mass gap. A possible microscopic route to such a term is discussed at the level of an effective high-frequency dressing construction in Appendix~\ref{app:floquet_derivation}.

The mass terms gap the Dirac cones and generate Berry curvature, while the scalar term shifts the effective Dirac sectors in opposite energy directions. The gray plane in Fig.~\ref{fig:proposal_schematic} denotes the chemical potential $\mu$. When $\mu$ intersects only part of the energy shifted massive Dirac spectrum, the occupied Berry curvature is dominated by the corresponding sector contribution. This selective occupation is the basic mechanism underlying the intermediate Hall response plateau discussed below.

The effective Bloch Hamiltonian is written as
\begin{equation}
    H(\mathbf{k})
    =
    d_0(\mathbf{k})\sigma_0
    +
    \mathbf{d}(\mathbf{k})\cdot\boldsymbol{\sigma},
    \label{eq:bloch_hamiltonian}
\end{equation} 
where $\sigma_0$ is the $2\times2$ identity matrix and $\boldsymbol{\sigma} (\sigma_x,\sigma_y,\sigma_z)$ are Pauli matrices acting on the two-component pseudospin degree of freedom of the $\pi$-flux unit cell. The crystal momentum is $\mathbf{k}=(k_x,k_y)$, with momenta measured in units of the inverse lattice spacing. 

The vector part of the Hamiltonian is
\begin{equation}
    \mathbf{d}(\mathbf{k})
    =
    \left(
    -2J\cos k_y,\,
    2m_T\sin k_x\sin k_y-m_I,\,
    -2J\cos k_x
    \right),
    \label{eq:d_vector}
\end{equation}
and the scalar part is
\begin{equation}
    d_0(\mathbf{k})
    =
    \Delta_v\sin k_x\sin k_y .
    \label{eq:d0_term}
\end{equation}
Here $J$ is the nearest-neighbor hopping amplitude and sets the energy scale. The coefficient $m_T$ controls the time reversal symmetry breaking Floquet mass, while $m_I$ denotes a static inversion breaking mass offset. The parameter $\Delta_v$ controls the strength of the momentum dependent scalar displacement. Because $d_0(\mathbf{k})$ multiplies $\sigma_0$, it changes the band energies but leaves the Bloch eigenvectors, and hence the Berry curvature, unchanged.

Unless otherwise stated, all energies are expressed in units of the nearest-neighbor hopping scale $J$. In the numerical calculations we set $J=1$, so that $\mu$, $m_T$, $m_I$, $\Delta_v$, and $E_\pm(\mathbf{k})$ are dimensionless when quoted in units of $J$. In an optical lattice realization, the microscopic energy scale is conventionally referenced to the recoil energy
\begin{equation}
    E_R=\frac{\hbar^2 k_L^2}{2M},
\end{equation}
where $M$ is the atomic mass and $k_L$ is the optical wave vector. The hopping amplitude $J$ is an effective tight-binding parameter determined by the lattice depth and laser-assisted tunneling matrix elements but the numerical plateau widths quoted in units of $J$ can be converted to recoil-energy units after specifying the microscopic ratio $J/E_R$ for a particular implementation.

\begin{figure*}[t]
    \centering
    \includegraphics[width=0.9\textwidth]{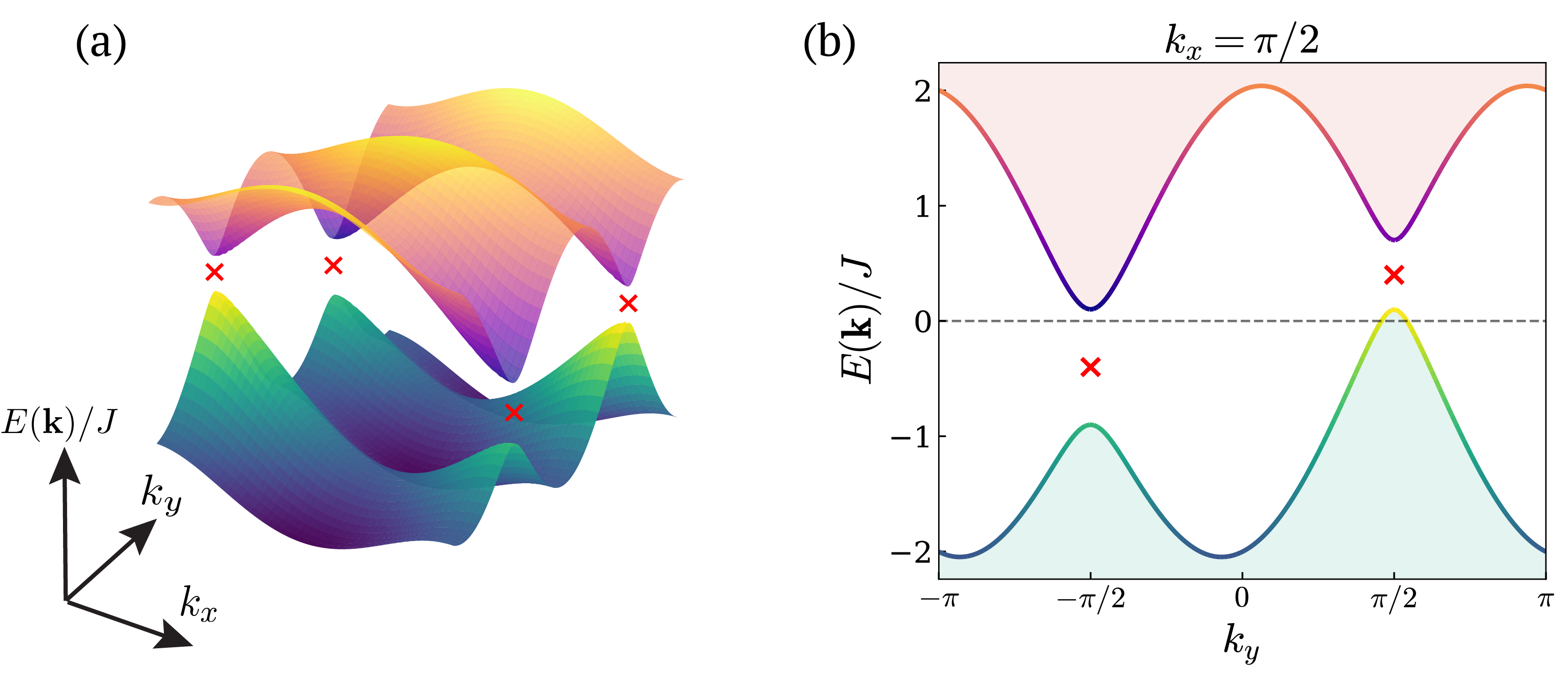}
    \caption{
    Energy spectrum of the effective $\pi$-flux lattice model. (a) Three dimensional band structure showing the massive Dirac cones, marked by red crosses, after the inclusion of the mass and scalar shift terms. (b) One dimensional cut of the spectrum at $k_x=\pi/2$ as a function of $k_y$, illustrating the gap opening and relative vertical displacement of the Dirac sectors.
    }
    \label{fig:band_structure}
\end{figure*}

The energy eigenvalues are
\begin{equation}
    E_\pm(\mathbf{k})
    =
    d_0(\mathbf{k})
    \pm
    |\mathbf{d}(\mathbf{k})|,
    \label{eq:band_energies}
\end{equation}
where
\begin{widetext}
\begin{equation}
    |\mathbf{d}(\mathbf{k})|
    =
    \sqrt{
    4J^2\cos^2 k_x
    +
    4J^2\cos^2 k_y
    +
    \left(
    2m_T\sin k_x\sin k_y-m_I
    \right)^2
    }.
    \label{eq:d_norm}
\end{equation}
\end{widetext}
The mass parameters therefore determine the Dirac gaps, while the scalar term $\Delta_v$ shifts the energies of the effective Dirac sectors relative to one another.

\section{Low energy Dirac expansion}
\label{sec:dirac_structure}

The model in Eq.~\eqref{eq:bloch_hamiltonian} is built from a $\pi$-flux kinetic term dressed by mass and scalar shift contributions. We now derive the corresponding low energy Dirac description, which will provide the basis for interpreting the Berry curvature and staircase response in later sections.

In the absence of the mass and scalar terms, $m_T=m_I=\Delta_v=0$, the spectrum becomes gapless when the kinetic components vanish,
\begin{equation}
    \cos k_x=0,
    \qquad
    \cos k_y=0.
\end{equation}
The four Dirac points are therefore located at
\begin{equation}
    \mathbf{K}_{s_xs_y}
    =
    \left(
    s_x\frac{\pi}{2},
    s_y\frac{\pi}{2}
    \right),
    \qquad
    s_x,s_y=\pm1,
    \label{eq:dirac_points}
\end{equation}
where the indices $s_x$ and $s_y$ label the signs of the corresponding Dirac-point momenta.

To obtain the continuum Hamiltonian near a given cone, we expand around
\begin{equation}
    \mathbf{k}
    =
    \mathbf{K}_{s_xs_y}
    +
    \mathbf{q},
    \qquad
    |\mathbf{q}|\ll1.
\end{equation}
Using
\begin{equation}
    \cos\left(
    s_x\frac{\pi}{2}+q_x
    \right)
    \simeq
    -s_xq_x,
    \qquad
    \cos\left(
    s_y\frac{\pi}{2}+q_y
    \right)
    \simeq
    -s_yq_y,
\end{equation}
together with
\begin{equation}
    \sin\left(
    s_x\frac{\pi}{2}+q_x
    \right)
    \simeq
    s_x,
    \qquad
    \sin\left(
    s_y\frac{\pi}{2}+q_y
    \right)
    \simeq
    s_y,
\end{equation}
the Hamiltonian becomes, to leading order in $\mathbf q$,
\begin{equation}
    H_{s_xs_y}(\mathbf q)
    =
    \epsilon_{s_xs_y}\sigma_0
    +
    v_{y,s_y}q_y\sigma_x
    +
    M_{s_xs_y}\sigma_y
    +
    v_{x,s_x}q_x\sigma_z.
    \label{eq:dirac_expanded_hamiltonian}
\end{equation}
The assignment of the linear momentum terms follows directly from the convention chosen in Eq.~\eqref{eq:d_vector}: the $k_y$ kinetic term multiplies $\sigma_x$, whereas the $k_x$ kinetic term multiplies $\sigma_z$.

The corresponding Dirac velocities are
\begin{equation}
    v_{x,s_x}
    =
    2Js_x,
    \qquad
    v_{y,s_y}
    =
    2Js_y,
    \label{eq:dirac_velocities}
\end{equation}
while the cone-dependent Dirac masses and scalar energy shifts are
\begin{equation}
    M_{s_xs_y}
    =
    2m_Ts_xs_y-m_I,
    \label{eq:dirac_mass}
\end{equation}
and
\begin{equation}
    \epsilon_{s_xs_y}
    =
    \Delta_v s_xs_y.
    \label{eq:dirac_shift}
\end{equation}

In the absence of the mass terms and the scalar displacement, the low energy Hamiltonian near each Dirac cone preserves the effective parity symmetry and time reversal symmetry of a two dimensional massless Dirac theory. In the present convention the kinetic Dirac Hamiltonian
\begin{equation}
    H_D(\mathbf{q}) = v_{y,s_y}q_y\sigma_x + v_{x,s_x}q_x\sigma_z
\end{equation}
is invariant under the combined transformation
\begin{equation}
    \mathbf{q} \rightarrow -\mathbf{q}, \qquad \psi_\mathbf{k} \rightarrow \sigma_y \psi_\mathbf{k}
\end{equation}
which acts as an effective parity operation within the low energy Dirac subspace,
\begin{equation}
    UH_D(-\mathbf{q})U^{-1} = H_D(\mathbf{q})
\end{equation}
with $U$ being a $\pi-$rotation about the $y$-axis in the pseudospin space. The Floquet generated TRS breaking mass term and the inversion breaking mass term breaks this effective parity symmetry and generates Berry curvature localized near the gapped Dirac cones. Near an isolated massive Dirac cone, the concentrated Berry curvature around the gap opening integrates to a Hall contribution whose sign is determined by the product of the Dirac mass and the cone chirality. In the continuum Dirac theory, this produces the characteristic half-quantized anomalous Hall response associated with the parity anomaly. In the present lattice system, the anomalous Hall response is obtained from the Berry curvature of the full band structure which can separate into two effective Dirac sectors. Eqs.~\eqref{eq:dirac_mass} and \eqref{eq:dirac_shift} show that both the Dirac mass and scalar energy shift depend only on the product $s_xs_y$. The four Dirac cones therefore separate naturally into two following sectors,
\begin{equation}
    S_\pm:
    \qquad
    s_xs_y=\pm1.
    \label{eq:Dirac_sectors}
\end{equation}
For the $S_+$ sector,
\begin{equation}
    M_+
    =
    2m_T-m_I,
    \qquad
    \epsilon_+
    =
    +\Delta_v,
    \label{eq:plus_sector}
\end{equation}
whereas for the $S_-$ sector,
\begin{equation}
    M_-
    =
    -2m_T-m_I,
    \qquad
    \epsilon_-
    =
    -\Delta_v.
    \label{eq:minus_sector}
\end{equation}
The corresponding gap scales are set by $|M_+|$ and $|M_-|$, as shown in the schematic in Fig.~\ref{fig:proposal_schematic}.

\begin{figure*}
    \centering
    \includegraphics[width=1.0\textwidth]{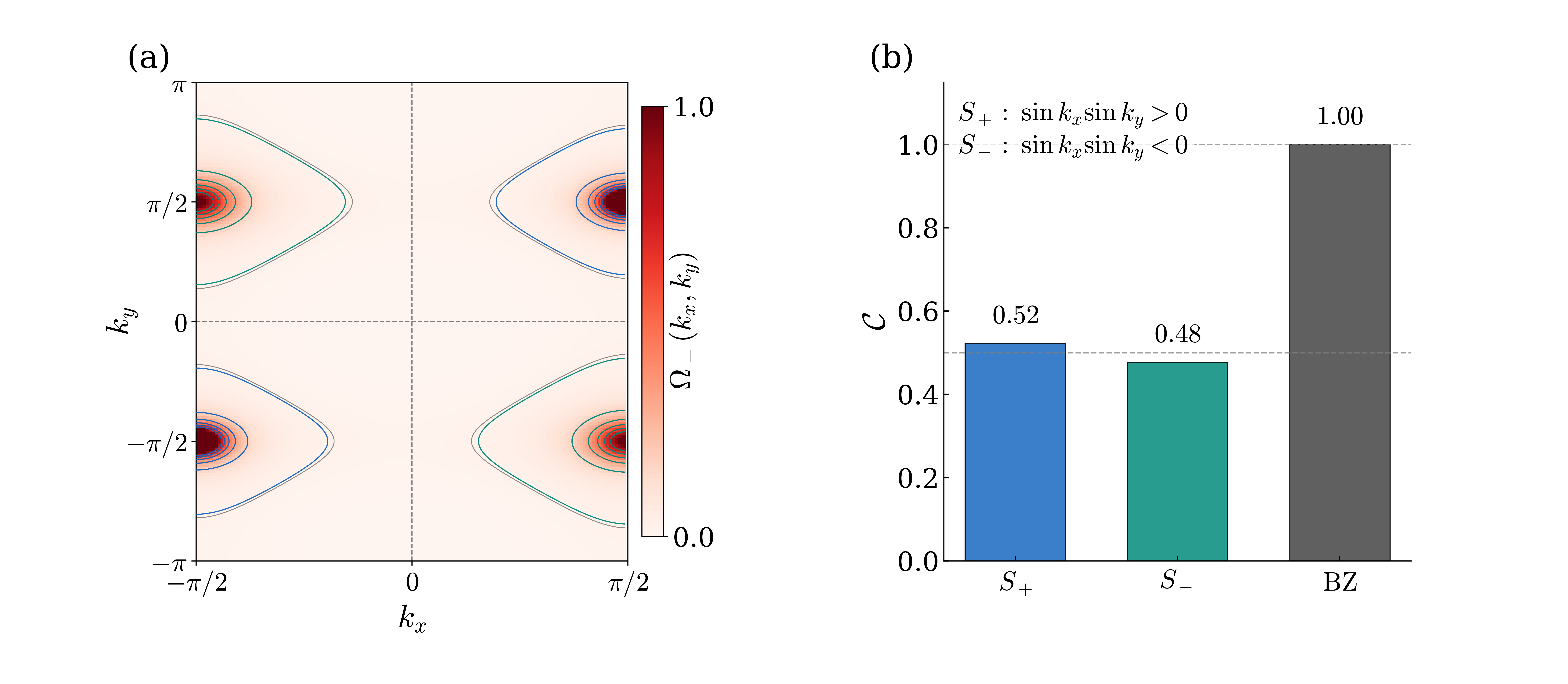}
    \caption{Berry curvature and sector resolved Berry curvature integrals of the effective two-band model.(a) Berry curvature $\Omega_-(\mathbf{k})$ of the lower band, with contours highlighting the localization of the curvature near the gapped Dirac cones. The signs of the peaks distinguish cones with different mass-chirality products, while the peak widths are controlled by the corresponding effective Dirac masses $M_\pm$. (b) Sector resolved Berry curvature integral $\mathcal{C}$ obtained by integrating $\Omega_-(\mathbf{k})$ over the corresponding Dirac sectors. The sector integrals show the half-integer Dirac contributions associated with the massive cones.
    }
    \label{fig:berry_curvature}
\end{figure*}

This decomposition separates the two ingredients underlying the staircase mechanism. The masses $M_\pm$ multiply $\sigma_y$ and therefore modify the Dirac eigenvectors, generating Berry curvature near the gapped cones. By contrast, the scalar shifts $\epsilon_\pm$ multiply $\sigma_0$ and therefore leave the eigenvectors unchanged while shifting the two sectors in opposite energy directions. The Berry curvature is therefore generated by the massive Dirac structure, whereas its contribution to the Hall response depends on whether the corresponding sector is occupied.

The four Dirac points in Eq.~\eqref{eq:dirac_points} are naturally grouped into two sectors according to the sign of $s_xs_y$. This grouping follows from a magnetic Bloch symmetry of the $\pi$-flux Hamiltonian \cite{ShaferPRB2023}. Defining $\mathbf Q=(\pi,\pi)$, the Bloch Hamiltonian satisfies
\begin{equation}
H(\mathbf k+\mathbf Q)
=
\sigma_y H(\mathbf k)\sigma_y .
\end{equation}
Consequently, Dirac points related by $\mathbf Q$ have identical spectra and equivalent low-energy structure. In particular, $(\pi/2,\pi/2)$ is related to $(-\pi/2,-\pi/2)$, while $(\pi/2,-\pi/2)$ is related to $(-\pi/2,\pi/2)$. The four extended zone Dirac points therefore organize into two symmetry related sectors, denoted $S_+$ and $S_-$ in Eq.~\ref{eq:Dirac_sectors}. Thus each sector $S_\pm$ appears as a single effective Dirac cone, rather than as two independent cones. The staircase response discussed later may be interpreted in terms of selective occupations and compensation of these two sector resolved Dirac contributions. We will use this effective sector viewpoint below when discussing the Berry curvature and the stepwise Hall response.

\section{Band structure and Berry curvature}
\label{sec:band_berry}

Having identified the low energy Dirac sectors of the effective Floquet Hamiltonian, we now examine how the mass and scalar shift terms appear in the band structure and Berry curvature. This provides the connection between the sector decomposition derived in Sec.~\ref{sec:dirac_structure} and the Hall response calculations presented below. The central observation is that the scalar term $d_0(\mathbf{k})$ shifts the quasiparticle energies without modifying the Bloch eigenvectors. Consequently, the Berry curvature is determined entirely by the vector part $\mathbf d(\mathbf k)$, whereas the occupation of that curvature depends on both the shifted band energies and the chemical potential.

The band energies are given by Eq.~\eqref{eq:band_energies}, and the corresponding spectrum is shown in Fig.~\ref{fig:band_structure}. Fig.~\ref{fig:band_structure}(a) displays the full energy surface, with the red crosses marking the locations of the massive Dirac cones. The mass parameters $m_T$ and $m_I$ open gaps at the Dirac points, producing the sector-dependent masses $M_\pm$ defined in Eqs.~\eqref{eq:plus_sector} and \eqref{eq:minus_sector}. The scalar shift $\Delta_v$ then displaces the two sectors in opposite energy directions: the $S_+$ sector is shifted by $+\Delta_v$, whereas the $S_-$ sector is shifted by $-\Delta_v$.

This separation is more clearly visible in Fig.~\ref{fig:band_structure}(b), which shows a one dimensional cut of the spectrum at fixed $k_x=\pi/2$ as a function of $k_y$. The cut highlights the two ingredients underlying the staircase mechanism: the mass terms gap the Dirac cones, while the scalar term shifts the corresponding sectors relative to one another in energy. As a result, different portions of the Berry curvature distribution become occupied as the chemical potential is varied.

For a generic two-level Bloch Hamiltonian, the Berry curvature of the lower band takes the standard form \cite{xiao_berry_2010}
\begin{equation}
    \Omega_-(\mathbf{k})
    =
    -\frac{1}{2}
    \hat{\mathbf d}(\mathbf{k})
    \cdot
    \left[
    \partial_{k_x}\hat{\mathbf d}(\mathbf{k})
    \times
    \partial_{k_y}\hat{\mathbf d}(\mathbf{k})
    \right],
    \label{eq:berry_curvature_lower}
\end{equation}
where
\begin{equation}
    \hat{\mathbf d}(\mathbf{k})
    =
    \frac{\mathbf d(\mathbf{k})}
    {|\mathbf d(\mathbf{k})|}.
\end{equation}
The Berry curvature of the upper band satisfies
\begin{equation}
    \Omega_+(\mathbf{k})
    =
    -\Omega_-(\mathbf{k}).
\end{equation}
The scalar term $d_0(\mathbf{k})$ does not appear in Eq.~\eqref{eq:berry_curvature_lower} because it multiplies the identity matrix and therefore leaves the Bloch eigenvectors unchanged. Its role is instead to determine which portions of the Berry curvature distribution are occupied at a given chemical potential.

Near a Dirac point labelled by $(s_x,s_y)$, the lattice Berry curvature reduces to the familiar massive-Dirac form. Using the linearized Hamiltonian in Eq.~\eqref{eq:dirac_expanded_hamiltonian}, the lower band curvature near the corresponding cone is
\begin{equation}
    \Omega_{-,s_xs_y}(\mathbf q)
    \simeq
    -\frac{1}{2}
    \frac{
    M_{s_xs_y}
    v_{x,s_x}
    v_{y,s_y}
    }{
    \left(
    v_{x,s_x}^2q_x^2
    +
    v_{y,s_y}^2q_y^2
    +
    M_{s_xs_y}^2
    \right)^{3/2}
    }.
    \label{eq:dirac_berry_curvature}
\end{equation}
This expression shows explicitly that the sign of the local Berry curvature is determined by the product of the Dirac mass and the cone chirality. In the present convention,
\begin{equation}
    v_{x,s_x}v_{y,s_y}
    =
    4J^2s_xs_y,
\end{equation}
so the chirality changes sign between the $S_+$ and $S_-$ sectors.

The full lattice Berry curvature is shown in Fig.~\ref{fig:berry_curvature}(a). In the parameter regime considered here, the dominant curvature weight is concentrated near the gapped Dirac cones, as emphasized by the contour lines. The width of each curvature peak is controlled by the corresponding effective mass scale: smaller $|M_\pm|$ produces a sharper and more localized curvature profile, whereas larger $|M_\pm|$ spreads the curvature over a broader region of momentum space. This behavior is consistent with the Dirac expression in Eq.~\eqref{eq:dirac_berry_curvature}.

To characterize the relative contribution of the two sectors, we define sector-resolved Berry curvature integrals
\begin{equation}
    \mathcal{C}_{S_\pm}
    =
    \frac{1}{2\pi}
    \int_{S_\pm}
    d^2k\,
    \Omega_-(\mathbf{k}),
    \label{eq:sector_chern}
\end{equation}
where the integration regions are chosen to isolate the momentum-space neighborhoods associated predominantly with the corresponding Dirac sectors. The sector decomposition used above reflects the same magnetic-Bloch structure discussed in Sec.~\ref{sec:dirac_structure}. Although the extended zone description contains four Dirac points, the pairs related by $\mathbf Q=(\pi,\pi)$ are not independent, they possess identical local spectra and Berry-curvature profiles. As a result, the Berry curvature organizes naturally into two sector-resolved weights associated with $S_+$ and $S_-$. The staircase mechanism is most conveniently interpreted in terms of the occupation and compensation of these two sector contributions.

The resulting sector-resolved Berry curvature integrals are shown in Fig.~\ref{fig:berry_curvature}(b). In the continuum massive-Dirac limit, each sector contributes a Berry curvature weight approaching a half-integer value, with the sign determined by the corresponding Dirac mass and cone chirality. In the full lattice model, these sector-resolved contributions provide a useful framework for interpreting the partial Hall response. The staircase structure arises when the chemical potential, or equivalently the scalar shift, selectively occupies the Berry curvature weight associated predominantly with one shifted massive Dirac sector before the other sector becomes significantly occupied.

\begin{figure}[t]
    \centering
    \includegraphics[width=0.9\linewidth]{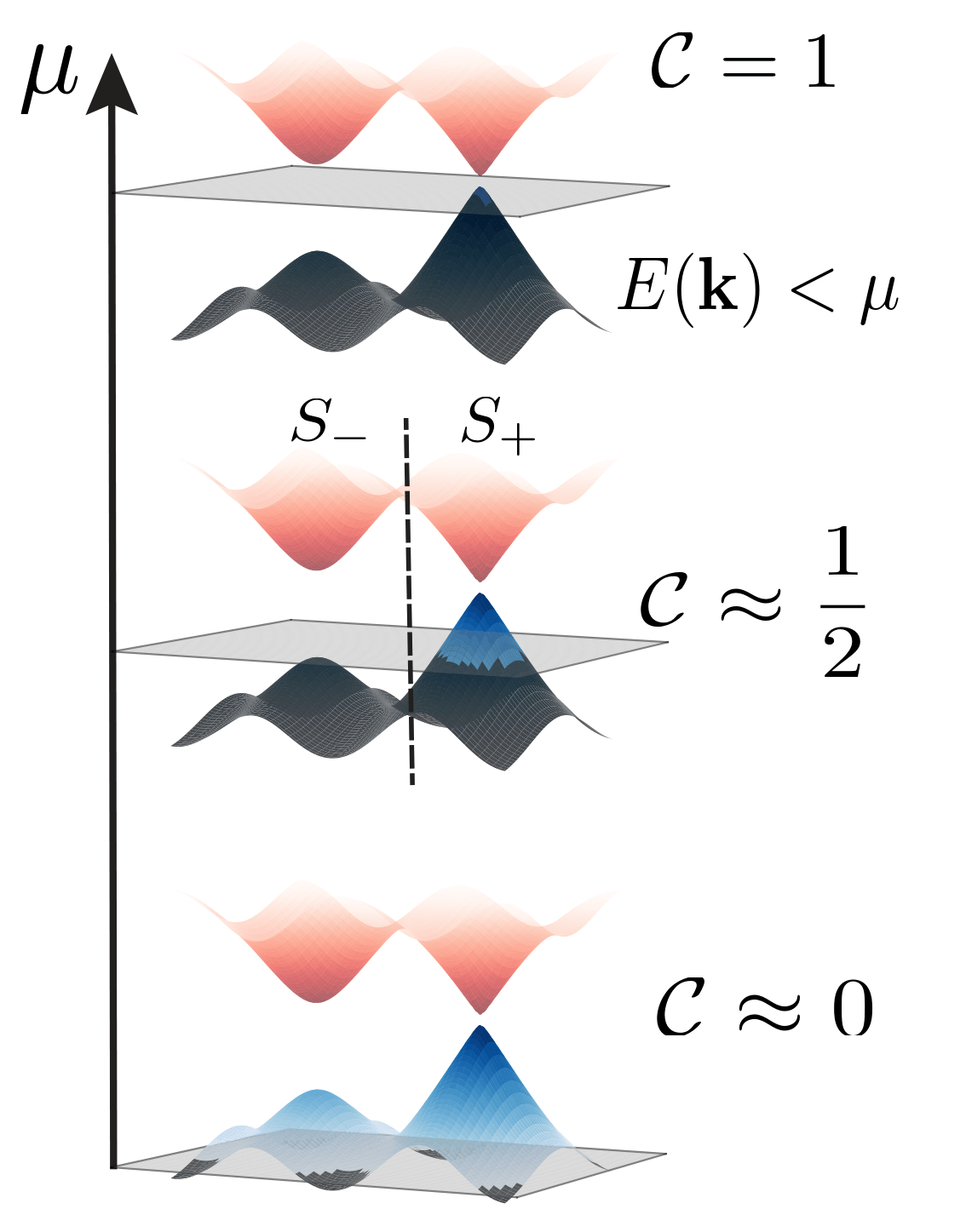}
    \caption{Schematic of the chemical potential-induced Hall staircase. The massive Dirac sectors $S_+$ and $S_-$ are shifted in opposite energy directions by the scalar term $\Delta_v\sin k_x\sin k_y$, while the mass parameters $m_T$ and $m_I$ open sector-dependent gaps. Each sector is represented by one effective massive Dirac cone; half-cone pieces appearing at the Brillouin-zone boundary are identified by periodic boundary conditions as a single effective Dirac contribution. The gray plane denotes the chemical potential $\mu$. As $\mu$ is swept through the shifted spectrum, the occupied Berry curvature contribution changes from $\mathcal{C}\simeq0$ to an intermediate response near $\mathcal{C}\simeq1/2$, and then to the next plateau near $\mathcal{C}\simeq1$.
    }
    \label{fig:mu_staircase_schematic}
\end{figure}

\section{Hall Response from Chemical Potential Tuning}
\label{sec:mu_staircase}

Having established that the Berry curvature is concentrated near the gapped Dirac sectors, we now examine how the scalar energy separation between these sectors produces a staircase structure in the Hall response. The essential point is that the scalar term $d_0(\mathbf{k})$ shifts the $S_+$ and $S_-$ sectors in opposite energy directions without modifying their Berry curvature. Varying the chemical potential $\mu$ therefore changes which sector-resolved Berry curvature contributions are occupied.

\begin{figure*}[t]
    \centering
    \includegraphics[width=0.9\textwidth]{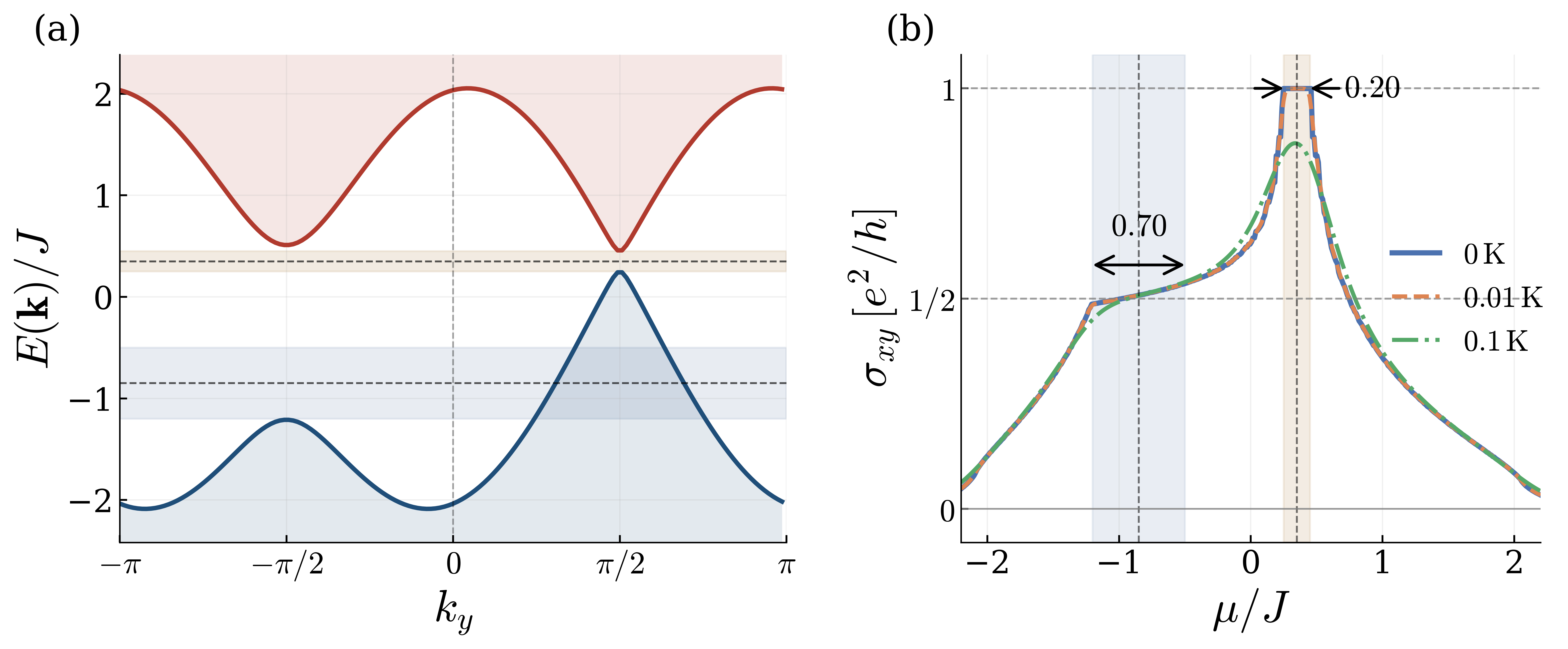}
    \caption{Chemical potential-induced anomalous Hall staircase. (a) One dimensional cut of the energy spectrum at $k_x=\pi/2$, with shaded regions indicating the chemical potential windows associated with the Hall plateaus. The blue region, centered near $E(\mathbf{k})/J=-0.8$, corresponds to occupation of one effective Dirac sector contribution and gives $\sigma_{xy}\simeq e^2/2h$. The yellow region, centered near $E(\mathbf{k})/J=0.3$, corresponds to occupation of both relevant Dirac sector contributions and gives $\sigma_{xy}\simeq e^2/h$. (b) Anomalous Hall response $\sigma_{xy}$ as a function of $\mu$, showing plateau-like regions generated by the sequential activation of Berry curvature contributions from the shifted massive Dirac sectors. The intermediate and upper plateaus have approximate widths $0.7$ and $0.2$, respectively.
    }
    \label{fig:sigma_xy_mu}
\end{figure*}

At some finite temperature, the anomalous Hall response of the partially occupied bands is computed from the Berry curvature weighted by the occupation function,
\begin{equation}
    \sigma_{xy}
    =
    \frac{1}{2\pi}
    \sum_{n=\pm}
    \int_{\mathrm{BZ}}
    d^2k\,
    f\!\left[
    E_n(\mathbf{k})-\mu
    \right]
    \Omega_n(\mathbf{k}),
    \label{eq:sigma_xy_mu}
\end{equation}
where $f(E-\mu)$ denotes the Fermi occupation factor. Throughout this work, $\sigma_{xy}$ denotes the anomalous Hall response expressed in units of $e^2/h$.

The chemical potential sweep is illustrated schematically in Fig.~\ref{fig:mu_staircase_schematic}. The two effective sectors $S_+$ and $S_-$ are represented by energy shifted massive Dirac cones, while the gray plane denotes the chemical potential $\mu$. As $\mu$ is raised through the spectrum, it intersects the shifted sectors sequentially. For sufficiently low $\mu$, neither sector contributes a substantial uncompensated Berry curvature weight, and the response remains near $\sigma_{xy}\simeq0$. When the chemical potential crosses the lower shifted sector, the occupied states become dominated by the Berry curvature associated primarily with one effective massive Dirac sector, producing an intermediate response near $\sigma_{xy}\simeq e^2/2h$. At larger $\mu$, the second shifted sector also becomes completely occupied, and the response approaches the next plateau near $\sigma_{xy}\simeq e^2/h$.

The intermediate plateau can be understood using the sector-resolved Berry curvature contributions discussed in Sec.~\ref{sec:band_berry}. In the continuum massive-Dirac limit, a single gapped Dirac cone contributes a Berry curvature weight approaching a half-integer value, with a sign determined by the Dirac mass and cone chirality. In the lattice model, however, the Hall response is always computed from the full Brillouin-zone integral in Eq.~\eqref{eq:sigma_xy_mu}. Because the Berry curvature remains strongly localized near the gapped Dirac sectors, the numerical staircase can nevertheless be interpreted in terms of the sequential occupation of the dominant sector-resolved Berry curvature weight.

The corresponding numerical results are shown in Fig.~\ref{fig:sigma_xy_mu}. Fig.~\ref{fig:sigma_xy_mu}(a) displays a one dimensional cut of the energy spectrum at fixed $k_x=\pi/2$, where the blue and yellow shaded regions indicate the chemical potential windows associated with the intermediate and upper plateaus, respectively. In the blue region, centered near $E(\mathbf{k})/J=-0.8$, the occupied states are dominated primarily by one shifted Dirac sector, producing a response near $\sigma_{xy}\simeq e^2/2h$. In the yellow region, centered near $E(\mathbf{k})/J=0.3$, both relevant sectors contribute substantially, producing a response near $\sigma_{xy}\simeq e^2/h$. The full dependence of $\sigma_{xy}$ on $\mu$ is shown in Fig.~\ref{fig:sigma_xy_mu}(b), where the same shaded regions identify plateau-like features with approximate widths $\delta\mu=0.7J$ and $0.2J$, respectively.

The chemical potential sweep therefore provides a direct probe of the staircase mechanism. The sector masses $M_\pm$ determine the Berry curvature generated near the gapped Dirac sectors, while the scalar shift $\Delta_v$ determines their relative energy separation. At fixed $\Delta_v$, varying $\mu$ moves the occupation threshold through the shifted sectors and produces the corresponding sequence of Hall response plateaus. The same interplay between Berry curvature generation and sector occupation can be probed in a complementary way by holding $\mu$ fixed and varying $\Delta_v$, as discussed in the next section.

\section{Hall Response from Scalar-Sector Displacement}
\label{sec:delta_staircase}

The preceding section considered a chemical potential sweep at fixed Dirac sector separation. We now turn to the complementary protocol in which the chemical potential is held fixed while the scalar shift $\Delta_v$ is varied. Instead of moving the occupation threshold through a fixed spectrum, this protocol shifts the $S_+$ and $S_-$ sectors relative to a fixed chemical potential. It therefore provides an alternative route to the same sequential occupation and partial compensation of the sector-resolved Berry curvature contributions.

The relevant dependence follows from the sector shifts defined in Eq.~\eqref{eq:dirac_shift}. Increasing $\Delta_v$ moves the $S_+$ sector upward in energy and the $S_-$ sector downward in energy. Because this term multiplies $\sigma_0$, it does not modify the Berry curvature generated by the mass terms. Its role is instead to determine which portions of the lower- and upper- band Berry curvature distributions are occupied at the chosen chemical potential.

At fixed $\mu$, the Hall response is computed as
\begin{equation}
    \sigma_{xy}(\Delta_v)
    =
    \frac{1}{2\pi}
    \sum_{n=\pm}
    \int_{\mathrm{BZ}}
    d^2k\,
    f\!\left[
    E_n(\mathbf{k};\Delta_v)-\mu
    \right]
    \Omega_n(\mathbf{k}),
    \label{eq:sigma_xy_delta}
\end{equation}
where the $\Delta_v$ dependence enters through the band energies,
\begin{equation}
    E_\pm(\mathbf{k};\Delta_v)
    =
    \Delta_v\sin k_x\sin k_y
    \pm
    |\mathbf d(\mathbf{k})|.
\end{equation}
Within the present effective description, the Berry curvature $\Omega_n(\mathbf{k})$ remains unchanged as $\Delta_v$ is varied because the scalar shift does not alter the Bloch eigenvectors. Additional microscopic couplings between the scalar dressing channel and the mass terms are neglected in the present treatment.

\begin{figure*}[t]
    \centering
    \includegraphics[width=0.9\textwidth]{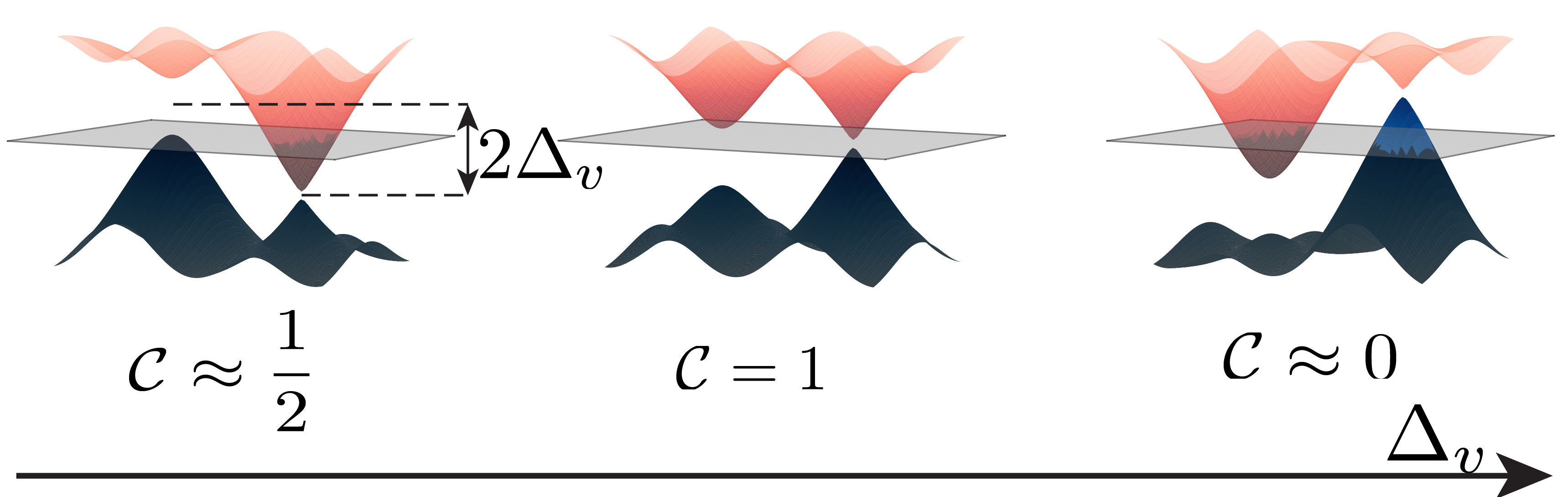}
    \caption{
    Schematic of the scalar shift-induced Hall staircase. The chemical potential $\mu$ is held fixed, shown by the gray plane, while the scalar shift $\Delta_v$ moves the massive Dirac sectors $S_+$ and $S_-$ in opposite energy directions. From left to right, increasing $\Delta_v$ changes the occupation of lower- and upper- band Berry curvature contributions. The three panels illustrate regimes with an uncompensated single sector response near $\mathcal{C}\simeq1/2$, a two sector response near $\mathcal{C}\simeq1$, and another compensated response returning toward $\mathcal{C}\simeq0$.
    }
    \label{fig:delta_staircase_schematic}
\end{figure*}

The scalar shift protocol is illustrated schematically in Fig.~\ref{fig:delta_staircase_schematic}. The gray plane denotes the fixed chemical potential. Unlike the chemical potential sweep, varying $\Delta_v$ can simultaneously change the occupation of both lower- and upper- band states at fixed $\mu$. Because the two bands carry opposite Berry curvature,
\begin{equation}
    \Omega_+(\mathbf{k})
    =
    -\Omega_-(\mathbf{k}),
\end{equation}
occupation of upper band states can partially compensate the contribution from the corresponding lower band sector. The staircase structure in $\Delta_v$ therefore arises from the combined effect of sequential sector occupation and partial cancellation between lower- and upper- band Berry curvature contributions.

As $\Delta_v$ is varied, the massive $S_+$ and $S_-$ sectors move in opposite energy directions relative to the fixed chemical potential plane. A sector begins to contribute when its lower band states move below $\mu$, while its contribution can become partially compensated once the corresponding upper band states also become occupied. This produces a staircase structure in $\sigma_{xy}$ analogous to the chemical potential sweep, but now controlled by the scalar sector displacement rather than by varying $\mu$ itself. In the first configuration of Fig.~\ref{fig:delta_staircase_schematic}, one sector contributes an uncompensated Berry curvature weight near $\sigma_{xy}\simeq e^2/2h$, while the other sector is largely compensated by occupied upper band states. At intermediate $\Delta_v$, both sectors contribute substantially, producing a response near $\sigma_{xy}\simeq e^2/h$. For larger $\Delta_v$, one sector is shifted above $\mu$ and becomes weakly occupied, while the remaining sector becomes increasingly compensated by upper band occupation, driving the net response back toward $\sigma_{xy}\simeq0$. 

\begin{figure}[h]
    \centering
    \includegraphics[width=\linewidth]{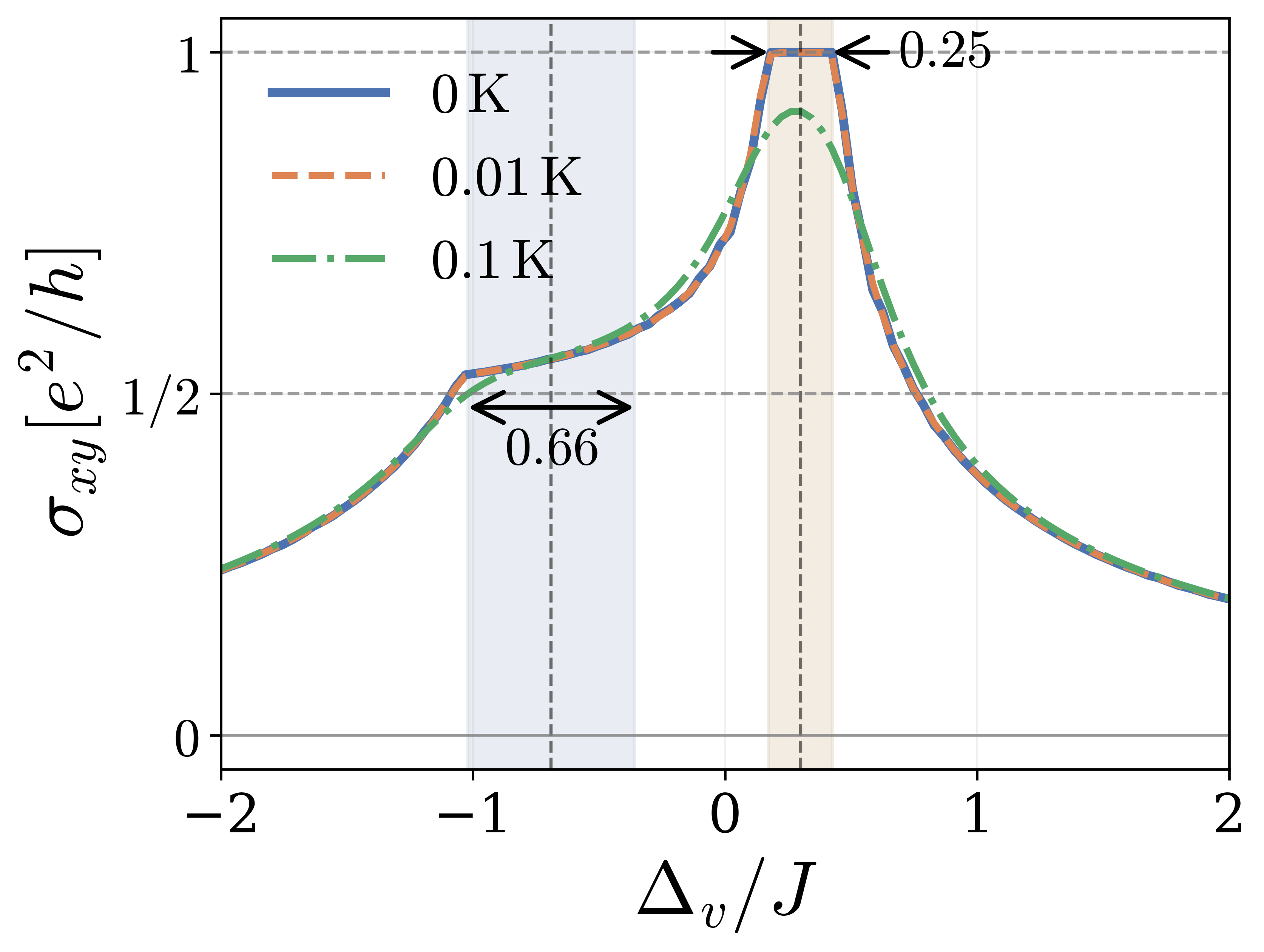}
    \caption{
    Scalar shift-induced anomalous Hall staircase at fixed chemical potential $\mu/J=0.3$. Varying the scalar shift $\Delta_v$ changes the relative energy displacement of the massive Dirac sectors $S_+$ and $S_-$, producing a staircase response through sequential sector activation and cancellation between lower- and upper- band Berry curvature contributions. The blue shaded region, centered near $\Delta_v/J=-0.89$, marks the intermediate plateau near $\sigma_{xy}\simeq e^2/2h$ with approximate width of $0.66$. The yellow shaded region, centered near $\Delta_v/J\simeq0.3$, marks the upper plateau near $\sigma_{xy}\simeq e^2/h$ with approximate width of $0.25$.
    }
    \label{fig:sigma_xy_delta}
\end{figure}

The numerically evaluated Hall response is shown in Fig.~\ref{fig:sigma_xy_delta}. Throughout this calculation, the chemical potential is fixed at $\mu/J=0.3$. The blue shaded region centered near $\Delta_v/J=-0.89$ indicates the intermediate plateau-like feature with approximate width $\delta\Delta_v=0.66J$, while the yellow shaded region centered near $\Delta_v/J\simeq0.3$ marks the higher response plateau-like feature with approximate width $\delta\Delta_v=0.25J$. The staircase structure again reflects the competition between sector occupation and compensation. In contrast to the chemical potential sweep, where the band structure is fixed and the occupation threshold is moved, the present protocol keeps $\mu$ fixed and instead shifts the sector energies themselves. The two protocols therefore probe the same underlying interplay between Berry curvature generation, sector displacement, and band occupation.

The scalar-shift protocol is experimentally appealing because it probes the staircase mechanism without requiring changes in the atomic filling or chemical potential. Instead, the relative displacement of the $S_+$ and $S_-$ sectors is controlled directly through the auxiliary scalar dressing channel. In the effective description developed here, the parameters $\Delta_v$, $m_T$, and $m_I$ are treated independently. In a microscopic implementation, however, their tunability may depend on the specific Floquet drive and AC-Stark dressing protocol used to realize the effective Hamiltonian.

\begin{figure*}[t]
    \centering
    \includegraphics[width=\textwidth]{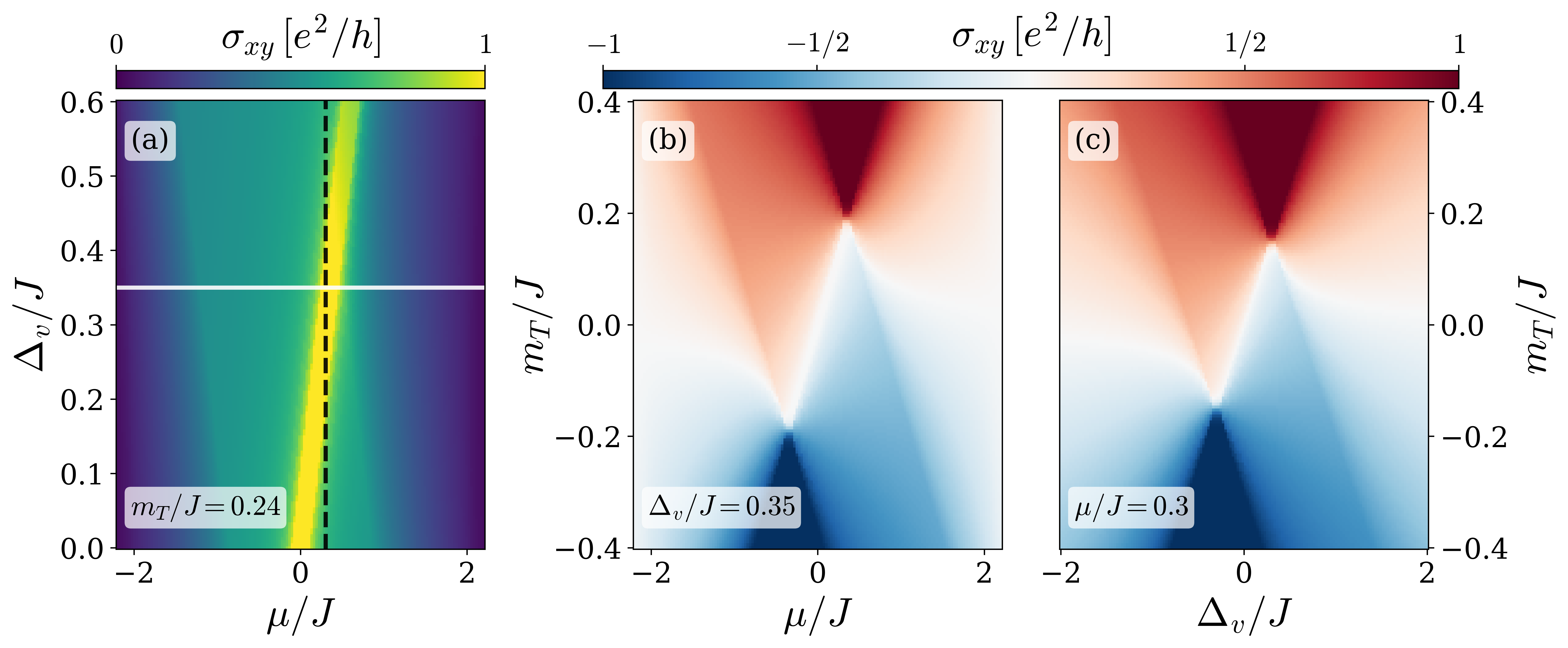}
    \caption{
    Hall response phase diagrams from the occupied Berry curvature integral of the full lattice model. The panels show $\sigma_{xy}$ in representative two parameter planes involving the chemical potential $\mu$, scalar shift $\Delta_v$, and time reversal symmetry breaking mass scale $m_T$. (a) Response in the $\mu$-$\Delta_v$ plane. The solid white and dashed black lines indicate the one dimensional cuts used in Figs.~\ref{fig:sigma_xy_mu} and \ref{fig:sigma_xy_delta}, respectively. (b) Response in the $\mu$-$m_T$ plane, showing the dependence of the chemical potential sweep on the sign and magnitude of the sector masses. (c) Response in the $\Delta_v$-$m_T$ plane, showing the corresponding dependence of the scalar shift sweep on $m_T$. Plateau-like regions arise from sequential activation and cancellation of Berry curvature contributions from the shifted massive Dirac sectors.
    }
    \label{fig:phase_diagrams}
\end{figure*}

\section{Hall response phase diagrams}
\label{sec:phase_diagrams}

The chemical potential and scalar shift sweeps discussed above correspond to one dimensional cuts through a larger parameter space. We now examine the Hall response as a function of multiple control parameters in order to identify the regions over which the plateau-like staircase features remain stable. This analysis also clarifies how the sector-selective response evolves as the Dirac masses and scalar energy shifts are varied simultaneously.

For each parameter set, the Hall response is computed from the occupied Berry curvature integral in Eq.~\ref{eq:sigma_xy_mu} evaluated using the full lattice Hamiltonian. The band energies depend on the scalar shift according to
\begin{equation}
    E_\pm(\mathbf{k})
    =
    \Delta_v\sin k_x\sin k_y
    \pm
    |\mathbf d(\mathbf{k})|,
\end{equation}
while the Berry curvature is determined by the vector part $\mathbf d(\mathbf{k})$ and therefore by the sector masses
\begin{equation}
    M_+
    =
    2m_T-m_I,
    \qquad
    M_-
    =
    -2m_T-m_I.
    \label{eq:sector_masses_phase}
\end{equation}
Consequently, $\mu$ and $\Delta_v$ primarily determine the occupation of the sector-resolved Berry curvature distributions, whereas $m_T$ determines the sign and magnitude of the corresponding massive-Dirac contributions.

The resulting Hall response is shown in Fig.~\ref{fig:phase_diagrams}. Fig.~\ref{fig:phase_diagrams}(a) displays $\sigma_{xy}$ in the $\mu$-$\Delta_v$ plane and combines the two staircase protocols discussed in the previous sections. The solid white and dashed black lines indicate the one dimensional cuts corresponding to the chemical potential and scalar shift sweeps, respectively. The extended plateau-like regions demonstrate that the intermediate and larger response regimes persist over finite parameter windows rather than appearing only at isolated fine tuned points. These regions occur when the chemical potential intersects different portions of the shifted massive sector spectrum, leading to changes in the occupation and compensation of the corresponding Berry curvature weight.

The structure of these diagrams can be understood from the low energy Dirac sector band edges. Near a Dirac point labelled by $(s_x,s_y)$, the band edges are approximately
\begin{equation}
    E_{\pm,s_xs_y}(\mathbf q=0)
    =
    \epsilon_{s_xs_y}
    \pm
    |M_{s_xs_y}|,
    \label{eq:dirac_band_edges}
\end{equation}
where $\epsilon_{s_xs_y}$ and $M_{s_xs_y}$ are defined in Eqs.~\eqref{eq:dirac_shift} and \eqref{eq:dirac_mass}. Approximate boundaries between different Hall response regions occur when $\mu$ crosses one of these sector-dependent band edges. Although the phase diagrams in Fig.~\ref{fig:phase_diagrams} are computed using the full lattice bands, Eq.~\eqref{eq:dirac_band_edges} provides a useful low energy guide to the observed staircase structure.

The $\mu$-$\Delta_v$ plane directly illustrates the interplay between the occupation threshold and the scalar sector displacement. Increasing $\mu$ moves the occupation threshold through a fixed spectrum, whereas changing $\Delta_v$ shifts the $S_+$ and $S_-$ sectors relative to a fixed threshold. Both operations can therefore drive transitions between the same Hall response regimes, as indicated by the one dimensional cuts in Fig.~\ref{fig:phase_diagrams}(a).

The planes involving $m_T$, shown in Figs.~\ref{fig:phase_diagrams}(b) and \ref{fig:phase_diagrams}(c), illustrate how the staircase structure depends on the time reversal symmetry breaking mass. For the parameter choice used in these panels, $m_I/J=0.38$, the sector masses are
\begin{equation}
    M_+
    =
    2m_T-0.38,
    \qquad
    M_-
    =
    -2m_T-0.38.
\end{equation}
For $m_T<m_I/2=0.19$, both sector masses are negative,
\[
M_+<0,
\qquad
M_-<0,
\]
so the two sectors share the same mass sign structure and the parity-anomaly inspired sector competition responsible for the staircase response is weak. At the critical value $2m_T\simeq m_I$, corresponding to $m_T/J\simeq0.19$, the $S_+$ sector becomes nearly gapless. In this regime the associated Berry curvature becomes strongly concentrated near the Dirac point and the staircase structure is less clearly resolved. For larger $m_T$, the two sectors acquire opposite mass signs and remain appreciably gapped, producing broader Berry curvature distributions and more robust plateau-like response regions.

The similarity between Figs.~\ref{fig:phase_diagrams}(b) and \ref{fig:phase_diagrams}(c) reflects the fact that the dominant qualitative evolution is controlled primarily by the sector masses $M_\pm(m_T)$. By contrast, varying $\mu$ or $\Delta_v$ mainly shifts the occupation boundaries between the same underlying response regimes. The phase diagrams therefore summarize the operating regime of the proposed staircase mechanism: robust intermediate response regions require both appreciable Dirac sector Berry curvature and sufficient scalar sector separation that the corresponding occupations can be distinguished independently.

\section{Conclusion}
\label{sec:conclusion}

We have proposed an effective cold atom realization of a parity anomaly inspired Hall staircase in a Floquet-engineered $\pi$-flux optical lattice. The starting point is a two-band lattice Hamiltonian containing four Dirac cones, which separate naturally into two effective sectors $S_+$ and $S_-$ due to magnetic Bloch symmetry of the $\pi$-flux lattice. An off-resonant circular Floquet drive generates a time reversal symmetry breaking mass, while a static inversion breaking offset provides an additional mass contribution. Together, these terms gap the two effective Dirac cones and generate Berry curvature localized near the massive Dirac sectors. An auxiliary scalar dressing channel produces a momentum dependent scalar shift that separates the  $S_+$ and $S_-$ sectors in energy without modifying their Bloch eigenvectors.

Using the occupied Berry curvature integral of the full lattice model, we showed that this sector separation produces a stepwise Hall response. Sweeping the chemical potential at fixed scalar shift changes the occupation of the shifted massive Dirac sectors, producing plateau-like response regions near $\sigma_{xy}\simeq e^2/2h$ and $\sigma_{xy}\simeq e^2/h$. We further showed that an analogous staircase structure can be obtained by holding the chemical potential fixed and varying the scalar shift $\Delta_v$, where the response is governed by both sequential sector occupation and partial compensation between lower- and upper- band Berry curvature contributions. The resulting two parameter phase diagrams demonstrate that these intermediate response regimes persist over finite regions of parameter space rather than occurring only at isolated fine tuned points. The predicted staircase response could be detected using established cold atom Hall probes based on force induced transverse center of mass motion, which directly reflects the Berry curvature distribution of the occupied bands.

The present work should be viewed as an effective-theory proposal rather than as a complete experimental blueprint. An important next step is to develop a more microscopic implementation of the auxiliary scalar dressing channel and to quantify the degree to which the parameters $m_T$, $m_I$, and $\Delta_v$ can be tuned independently in a realistic optical lattice setting. Further work should also address finite temperature effects, trap inhomogeneity, Floquet heating, and experimental protocols for extracting Hall response signatures in a neutral atom system. Nevertheless, the results presented here identify a concrete route by which Floquet and auxiliary dressing techniques can be combined to generate sector-selective occupation and compensation of massive Dirac sector Berry curvature, providing a cold atom platform for exploring parity anomaly inspired half-integer Hall responses in two dimensions.

\begin{acknowledgments}
We acknowledge support from ARO Grant No. ARO-W911NF2210247. S.T. and B.B.R acknowledge support from ONR Grant No. ONR-N000142312061. V.W.S. acknowledges support from Grant No. AFOSR-FA9550-23-1-0034. 
\end{acknowledgments}

\appendix

\section{Derivation of the effective Floquet Hamiltonian}
\label{app:floquet_derivation}

We begin from the static $\pi$-flux kinetic Hamiltonian
\begin{equation}
    H_0(\mathbf{k})
    =
    -2J\cos k_y\,\sigma_x
    -
    2J\cos k_x\,\sigma_z
    \label{eq:app_H0}
\end{equation}
This Hamiltonian contains the two kinetic components of the effective $\pi$-flux lattice and hosts Dirac points at
\begin{equation}
    \mathbf{K}_{s_xs_y}
    =
    \left(
    s_x\frac{\pi}{2},
    s_y\frac{\pi}{2}
    \right)
\end{equation}

To generate a time reversal symmetry breaking mass, we introduce an off-resonant periodic drive through the Peierls substitution
\begin{equation}
    \mathbf{k}\rightarrow\mathbf{k}+\mathbf{A}(t)
\end{equation}
with
\begin{equation}
    \mathbf{A}(t)
    =
    A_0(\cos\omega t,\eta\sin\omega t)
    \label{eq:app_drive}
\end{equation}
The driven Hamiltonian becomes
\begin{equation}
    H(\mathbf{k},t)
    =
    -2J\cos[k_y+A_y(t)]\,\sigma_x
    -
    2J\cos[k_x+A_x(t)]\,\sigma_z
    \label{eq:app_driven_H}
\end{equation}
For small drive amplitude, we expand to leading order in $A_0$:
\begin{align}
    H(\mathbf{k},t)
    &\simeq
    H_0(\mathbf{k})
    +
    2JA_y(t)\sin k_y\,\sigma_x
    +
    2JA_x(t)\sin k_x\,\sigma_z
    \label{eq:app_H_linear}
\end{align}

Writing the time-periodic Hamiltonian as
\begin{widetext}
\begin{equation}
    H(\mathbf{k},t)
    =
    \sum_n H_n(\mathbf{k})e^{in\omega t},
    \qquad
    H_n(\mathbf{k})
    =
    \frac{1}{T}
    \int_0^T dt\,
    H(\mathbf{k},t)e^{-in\omega t},
    \quad
    T=\frac{2\pi}{\omega},
\end{equation}
\end{widetext}
we obtain the first harmonics by using
\begin{equation}
    \cos\omega t
    =
    \frac{e^{i\omega t}+e^{-i\omega t}}{2},
    \qquad
    \sin\omega t
    =
    \frac{e^{i\omega t}-e^{-i\omega t}}{2i}.
\end{equation}
Substituting these expressions into Eq.~\eqref{eq:app_H_linear} and
collecting the coefficients of $e^{\pm i\omega t}$ gives
\begin{equation}
    H_{+1}
    =
    -i\eta JA_0\sin k_y\,\sigma_x
    +
    JA_0\sin k_x\,\sigma_z,
    \label{eq:app_H_plus}
\end{equation}
and
\begin{equation}
    H_{-1}
    =
    i\eta JA_0\sin k_y\,\sigma_x
    +
    JA_0\sin k_x\,\sigma_z .
    \label{eq:app_H_minus}
\end{equation}

In the off-resonant regime, the leading van Vleck effective Hamiltonian \cite{Goldman2014, eckardt_colloquium_2017, mikami_brillouin-wigner_2016} is
\begin{equation}
    H_{\mathrm{eff}}
    =
    H_0
    +
    \frac{[H_{-1},H_{+1}]}{\hbar\omega}
    +
    O(\omega^{-2})
    \label{eq:app_van_vleck}
\end{equation}
Using Eqs.~\eqref{eq:app_H_plus} and \eqref{eq:app_H_minus}, the
commutator gives
\begin{equation}
    [H_{-1},H_{+1}]
    =
    4\eta (JA_0)^2
    \sin k_x\sin k_y\,\sigma_y .
    \label{eq:app_commutator}
\end{equation}
Thus the off-resonant circular drive generates the effective mass term
\begin{equation}
    H_T(\mathbf{k})
    =
    2m_T\sin k_x\sin k_y\,\sigma_y 
    \label{eq:app_HT}
\end{equation}
with
\begin{equation}
    m_T
    =
    \frac{2\eta (JA_0)^2}{\hbar\omega}.
    \label{eq:app_mT}
\end{equation}
The sign of $m_T$ is controlled by the handedness of the drive through
$\eta$.

The inversion breaking mass offset is included as a static control term,
\begin{equation}
    H_I
    =
    -m_I\sigma_y
    \label{eq:app_HI}
\end{equation}
Microscopically, the inversion breaking mass offset may be generated through a sublattice energy imbalance, differential light shift, or Raman detuning that distinguishes the two pseudospin components of the $\pi$-flux unit cell.

The scalar shift used in the main text is modeled through an auxiliary off-resonant scalar AC Stark or Raman dressing channel \cite{SpielmanPRA2009,cooper_topological_2019,grimm_optical_2000}. Unlike the Peierls drive discussed above, this dressing couples symmetrically to the two pseudospin components and therefore produces a scalar quasienergy correction proportional to the identity matrix. We consider an auxiliary scalar modulation of the form
\begin{equation}
    V_{\mathrm{aux}}(\mathbf r,t)
    =
    V_0
    \cos(\omega_{\mathrm{aux}}t+\varphi)\,
    \sin(k_Lx)\sin(k_Ly)
    \label{eq:app_Vaux_real_space}
\end{equation}
where $V_0$ is the modulation amplitude, $\omega_{\mathrm{aux}}$ is the auxiliary driving frequency, $\varphi$ is a controllable phase offset, and $k_L$ is the wave vector associated with the auxiliary dressing beams. The spatial profile in Eq.~\eqref{eq:app_Vaux_real_space} is chosen such that, after projection onto the low energy Bloch basis of the $\pi$-flux lattice, the resulting coupling acquires the momentum structure required for the scalar shift term.

To describe this process, we introduce an auxiliary excited dressed state $\ket{e}$, detuned by $\Delta_e$ from the low energy manifold. In the rotating frame, the auxiliary Hamiltonian may be written schematically as
\begin{equation}
    H_{\mathrm{aux}}
    =
    \Delta_e\ket{e}\bra{e}
    +
    \left[
    g(\mathbf{k},t)\ket{e}\bra{g}
    +
    g^*(\mathbf{k},t)\ket{g}\bra{e}
    \right]
    \label{eq:app_aux_hamiltonian}
\end{equation}
where $\ket{g}$ denotes a state in the low energy two-band manifold. The effective momentum dependent coupling is obtained by projecting the auxiliary modulation onto the Bloch states,
\begin{equation}
    g(\mathbf{k},t)
    =
    \braket{e|V_{\mathrm{aux}}(\mathbf r,t)|u_{\mathbf{k}}}
    \label{eq:app_g_projection}
\end{equation}
with $\ket{u_{\mathbf{k}}}$ the low energy Bloch state at momentum $\mathbf{k}$.

The precise form of this matrix element depends on the auxiliary beam geometry and on the Wannier-orbital structure of the $\pi$-flux lattice. For the purposes of the effective model, we assume that the auxiliary dressing is chosen such that its projection onto the low-energy manifold has odd form factors along the two lattice directions. Keeping the leading harmonics compatible with this symmetry, the projected coupling may be parametrized as
\begin{equation}
    g(\mathbf{k},t)
    =
    \left[
    g_x\sin k_x
    +
    g_y e^{i\phi}\sin k_y
    \right]
    \cos(\omega_{\mathrm{aux}}t+\varphi),
    \label{eq:app_aux_coupling}
\end{equation}
where $g_x$ and $g_y$ are effective dressing amplitudes and $\phi$ is the relative interference phase between the two projected channels.

For large detuning, $|g|/|\Delta_e|\ll1$, the auxiliary state is only virtually occupied and may be eliminated perturbatively. Equivalently, second-order degenerate perturbation theory in the low-energy manifold gives an energy correction
\begin{equation}
    \delta E_g(\mathbf{k},t)
    \simeq
    \frac{
    |\bra{e}H_{\mathrm{aux}}\ket{g}|^2
    }{
    E_g-E_e
    }
    =
    -
    \frac{|g(\mathbf{k},t)|^2}{\Delta_e}.
\end{equation}
Because the auxiliary dressing is taken to couple approximately equally to the two pseudospin components, this shift is proportional to the identity operator in the low energy two-band subspace:
\begin{equation}
    \delta H_{\mathrm{Stark}}(\mathbf{k},t)
    \simeq
    -
    \frac{|g(\mathbf{k},t)|^2}{\Delta_e}\,\sigma_0 .
    \label{eq:app_stark_shift_general}
\end{equation}

To obtain the momentum structure used in the effective model, we use two interfering dressing channels with form factors proportional to $\sin k_x$ and $\sin k_y$:
\begin{equation}
    g(\mathbf{k},t)
    =
    \left[
    g_x\sin k_x
    +
    g_y e^{i\phi}\sin k_y
    \right]
    F(t)
    \label{eq:app_aux_coupling_interference}
\end{equation}
where $F(t)$ is a periodic modulation envelope. Substituting Eq.~\eqref{eq:app_aux_coupling_interference} into Eq.~\eqref{eq:app_stark_shift_general} and averaging over the fast modulation gives

\begin{widetext}
\begin{align}
    \delta H_{\mathrm{Stark}}(\mathbf{k})
    =
    -
    \frac{\overline{|F(t)|^2}}{\Delta_e}
    \big[
    g_x^2\sin^2 k_x
    +
    g_y^2\sin^2 k_y
    +
    2g_xg_y\cos\phi\,\sin k_x\sin k_y
    \big]\sigma_0 
    \label{eq:app_stark_expanded}
\end{align}
\end{widetext}
The first two terms renormalize the scalar dispersion and may be absorbed into the background band structure or compensated by weak static lattice potentials. Retaining the symmetry-relevant interference term gives the effective scalar contribution
\begin{equation}
    H_v(\mathbf{k})
    =
    \Delta_v\sin k_x\sin k_y\,\sigma_0
    \label{eq:app_Hv}
\end{equation}
with
\begin{equation}
    \Delta_v
    =
    -
    \frac{2g_xg_y\cos\phi}{\Delta_e}
    \overline{|F(t)|^2}
    \label{eq:app_Deltav}
\end{equation}
The sign and magnitude of $\Delta_v$ are therefore tunable through the relative phase $\phi$, the dressing amplitudes $g_x,g_y$, and the detuning $\Delta_e$.

Combining the static kinetic Hamiltonian, the Floquet-generated time reversal breaking mass, the inversion breaking offset, and the scalar AC-Stark dressing gives the effective Hamiltonian used in the main text:

\begin{widetext}
\begin{equation}
    H_F(\mathbf{k})
    =
    \Delta_v\sin k_x\sin k_y\,\sigma_0
    -
    2J\cos k_y\,\sigma_x
    +
    \left(
    2m_T\sin k_x\sin k_y
    -
    m_I
    \right)\sigma_y
    -
    2J\cos k_x\,\sigma_z
    \label{eq:app_final_HF}
\end{equation}
\end{widetext}
The mass terms modify the Bloch eigenvectors and generate Berry curvature, whereas the scalar term shifts the quasienergies of the $S_+$ and $S_-$ sectors without changing the eigenvectors.

\bibliographystyle{apsrev4-2}
\bibliography{Bibliography}

\end{document}